# Quasi-phasematched laser wakefield acceleration

S.J. Yoon, J.P. Palastro, and H.M. Milchberg

*Institute for Research in Electronics and Applied Physics, University of Maryland
College Park, Maryland 20740*

**Abstract**

The energy gain in laser wakefield acceleration (LWFA) is ultimately limited by dephasing, occurring when accelerated electrons outrun the accelerating phase of the wakefield. We apply quasi-phasematching, enabled by axially modulated plasma channels, to overcome this limitation. By matching the modulation period to the dephasing length, a relativistic electron can undergo energy gain over several dephasing lengths. Theory and simulations are presented showing that at a weakly relativistic laser intensity, ~$10^{17}$ W/cm$^2$, and millijoule level pulse energies, quasi-phasematched LWFA results in energy gains of 50 MeV larger than standard LWFA.

The ponderomotive force of a high intensity, ultrashort laser pulse displaces plasma electrons, exciting plasma waves [1]. Realizing that the axial electric field associated with the plasma wave far surpasses that of conventional accelerators, Tajima and Dawson suggested harnessing the wakefield for electron acceleration [2]. Since their seminal paper, the promise of building smaller-scale, cheaper 'advanced accelerators' has led to a number of theoretical advances [3-5] and experimental demonstrations [6-9] of laser wakefield acceleration (LWFA). Several groups have recently achieved a LWFA milestone: 2 GeV energy gain for electrons accelerated over multiple cm [10,11]. Without multiple stages, the energy gain in these experiments is ultimately limited by electrons outrunning the accelerating phase of the wakefield or dephasing.

The operating paradigm for recent nonlinear wakefield acceleration experiments is to set the distance over which the laser pulse energy is depleted by driving plasma waves to the dephasing length, the distance over which the accelerated electrons outrun the accelerating phase of the wakefield [10,12]. Because the dephasing length scales as

$L_d \propto n_0^{-3/2}$, where $n_0$ is plasma density, and the maximum axial field is limited to the plasma wave breaking field, $E_{max} \propto n_0^{1/2}$, lower densities increase the maximum energy gain of electrons, $\Delta\gamma \propto n_0^{-1}$ [5]. The laser pulse must also stay collimated over the dephasing length. Self-guiding, where the transverse ponderomotive force of the laser pulse bores a guiding structure in the plasma, offers one method for collimation, but requires higher pulse powers, $P_{sf}$ for relativistic self-focusing, as the density decreases, $P_{sf} \propto n_0^{-1}$ [5,12]. Alternatively, preformed plasma waveguides [7,13,14] can be used, eliminating the density and power dependence of the guiding condition.

Axial density modulated plasma waveguides have been developed [15] and explored for quasi-phasematched direct laser acceleration of electrons [16-19] and THz generation [20]. In an axially modulated plasma waveguide, the guided mode is composed of spatial harmonics whose associated phase velocities can be tuned through the modulation period. Quasi-phasematching (QPM) refers to matching the phase velocity of an individual spatial harmonic to the electron velocity [19]. Analogous to QPM in nonlinear optics, the periodic medium compensates the difference between the electron velocity and the phase velocity of the accelerating field to transfer energy from the accelerating field to the electron.

Here we investigate the application of QPM in modulated plasma channels to LWFA (QPM-LWFA). The frequency of the excited plasma wave, and consequently its phase velocity, undergoes oscillations in the modulated plasma. As a result, the plasma wave itself is composed of spatial harmonics. We will see that by matching the modulation period to the dephasing length, a relativistic electron can undergo energy gain over several dephasing lengths. Furthermore, QPM-LWFA can operate at much lower pulse energies and provides a guiding structure for the laser pulse, thus loosening the three energy gain limitations associated with standard LWFA: dephasing, depletion, and diffraction.

We start by examining the electrostatic fields of ponderomotively driven plasma waves in a corrugated plasma channel. The density profile of the plasma can be modeled as $n_e(r,z) = n_0[1 + \delta\sin(k_m z)] + \frac{1}{2}n_0'' r^2$, where $n_0$ is the averaged on axis density, $\delta$ is the modulation amplitude, $k_m = 2\pi/\lambda_m$ is the wavevector associated with the density modulations of period $\lambda_m$, and $n_0''$ describes the curvature of the channel. The transverse parabolic density profile provides guiding for a laser pulse with a exp(−1) field width $w_{ch} = (2c)^{1/2}(m_e/2\pi e^2 n_0'')^{1/4}$, where $c$ is the speed of light in vacuum, and $m_e$ and $e$ are the electron rest mass and charge respectively.

To illustrate the concept and to derive a scaling for the energy gain, we consider a weakly relativistic laser pulse propagating along the z-axis with wavelength $\lambda_0 = 2\pi/k_0$ and normalized vector potential, $\mathbf{a} = e\mathbf{A}/m_e c$, satisfying $|\mathbf{a}| < 0.5$. As the laser pulse propagates through the plasma, its ponderomotive force drives an electron plasma wave, a travelling wave with a phase velocity equal to the group velocity of the laser pulse. Using a separation of time scales based on the disparity between the laser pulse and plasma frequencies, the equation for the wakefield in a non-uniform plasma can be found from the fluid and Maxwell's equations:

$$\left[\frac{\partial^2}{\partial \xi^2} + k_p^2(r,z)\right]\mathbf{E} = -\pi e n_e(r,z)\nabla |\mathbf{a}|^2 \quad (1)$$

where $k_p^2 = \omega_p^2/c^2 = 4\pi e^2 n_e / m_e c^2$ and $\xi = z - v_g t$ is the coordinate measuring distance in a frame moving with the group velocity, $v_g$, of the laser pulse. We are primarily interested in the case of $\delta \ll 1$ such that $v_g$ is essentially constant, namely $v_g/c \simeq 1 - (k_p^2/2k_0^2) - (4/k_0^2 w_{ch}^2)$ where $k_{p0}^2 = \omega_{p0}^2/c = 4\pi e^2 n_0/m_e c^2$. Equation (1) demonstrates that the on-axis wavenumber and plasma period undergo periodic expansions and contractions as the density oscillates, $\lambda_p(0,z) = \lambda_{p0}[1-\delta\sin(k_m z)]^{-1/2}$.

We assume a laser pulse of the form $|\mathbf{a}(\xi,r)|^2 = a_0^2 \exp(-2r^2/w_{ch}^2)\sin^2(\pi\xi/c\sigma)$ on the domain $0 < \xi < c\sigma$ with temporal full width half maximum (FWHM) $\sigma_{FWHM} = \sigma/2$ matched to the on-axis plasma period, $\sigma_{FWHM} = \pi/\omega_{p0}$. For $\delta \ll 1$, the wakefields close to the axis, $r^2 \ll w_{ch}^2$, and after the laser pulse, $\xi > c\sigma$, are then

$$E_z = -\frac{\pi}{8}a_0^2 \sum_n J_n\left[\frac{\delta k_{p0}(v_g t - z)}{2}\right]\cos\left[k_{p0}v_g t - (nk_m + k_{p0})z\right] \quad (2a)$$

$$E_r = -\frac{a_0^2}{2k_{p0}w_{ch}}\left(\frac{r}{w_{ch}}\right)\sum_n J_n\left[\frac{\delta k_{p0}(v_g t - z)}{2}\right]\sin\left[k_{p0}v_g t - (nk_m + k_{p0})z\right] \quad (2b)$$

where the fields have been normalized to the wave breaking field, $m_e c\omega_{p0}/e$. Equation (2) exhibits the decomposition of the wakefields into spatial harmonics whose amplitudes depend on the distance behind the head of the laser pulse, and whose phase velocities depend on the modulation period.

Figure 1(a) shows the on-axis longitudinal electric field, $E_z$, of a plasma wave driven by a low-amplitude, $|\mathbf{a}|\ll 1$, $\lambda_0 = 800$ nm laser pulse as a function of $z - ct$ and $z$ in a corrugated plasma channel with $n_0 = 7\times 10^{18}$ cm$^{-3}$, $\delta = 0.04$, $w_{ch} = 15$ $\mu m$, and $\lambda_m = 5.0$ mm. The pulse duration and spot are matched to the density and channel curvature respectively. The red dashed line marks the path taken by the on-axis peak of the laser pulse. The pulse slides back in the speed of light frame on account of its subluminal group velocity. In a uniform channel, the phase fronts of the longitudinal wakefield are parallel to the pulse's trajectory in the $z - ct$ and $z$ plane, the red dashed line. In a corrugated channel, the pulse passes through oscillating plasma density, causing the wake phases to oscillate with respect to the pulse's trajectory.

From Eq. (2), the phase velocity, $v_{p,n}$, of the wakefield's $n^{th}$ spatial harmonic is

$$\frac{v_{p,n}}{c} \simeq 1 - \frac{k_{p0}^2}{2k_0^2} - \frac{4}{k_0^2 w_{ch}^2} - n\frac{k_m}{k_{p0}}. \quad (3)$$

The phase of the $n^{th}$ spatial harmonic can be made stationary in the speed of light frame by setting the modulation period to $\lambda_m = -2n\lambda_{p0}^3\lambda_0^{-2}(1+8/k_{p0}^2 w_{ch}^2)^{-1}$ where $\lambda_{p0} = 2\pi/k_{p0}$. For the $n = -1$ spatial harmonic, this is equivalent to setting the modulation period equal to the dephasing length, $L_d = 2\lambda_{p0}^3\lambda_0^{-2}(1+8/k_{p0}^2 w_{ch}^2)^{-1}$, of standard LWFA. When $\lambda_m = L_d$ an electron moving near the speed of light along the z-axis experiences a near constant axial acceleration from the $n = -1$ spatial harmonic, while the acceleration of all other spatial harmonics time averages to zero.

Figure 1(b) displays a line out of Fig. 1(a), in which $\lambda_m = L_d$, along the white dashed line in red. For comparison, a line out at the same point in a uniform plasma channel is displayed in black. These curves reveal the longitudinal field acting on an electron moving at nearly c. In both cases the axial field oscillates at the plasma period, but the oscillations in the modulated channel are clearly non-sinusoidal. While the integral of the axial field over a plasma period is zero in the uniform channel, it is non-zero in the modulated channel. This clearly shows that the modulated longitudinal wakefield performs non-zero work on a relativistic electron even after the electron has traversed a full plasma wavelength.

In Fig. 1(c) the phase space of axial momentum, $P_z$, and speed of light frame coordinate is plotted for a long, uniform beam of test electrons with initial axial momentum of 100 $MeV/c$ accelerated over 2 cm. The results were obtained from 2D particle-in-cell simulations which we discuss further below. The pulse amplitude, wavelength, and FWHM were $a_0 = 0.25$, $\lambda_0 = 800$ $nm$, and $\sigma_{FWHM} = 30$ $fs$ respectively, and the density parameters were the same as given above. We note that the FWHM is longer than matched pulse duration of $\pi/\omega_{p0} = 21$ $fs$. The red curve indicates the location of the laser pulse propagating to the right in the figure and the white curve indicates the normalized amplitude of the $n = -1$ spatial harmonic, $J_{-1}$. The plot clearly shows that axial momentum gain is proportional to the amplitude of the phase matched spatial harmonic. The spikes in momentum result from bunching of the positively accelerated electrons in each half-period of the plasma wave.

The energy gain of a relativistic electron accelerated by the phase matched harmonic can be found by integrating $d\gamma/dt = -\omega_{p0}(v_z/c)E_z$, where $\gamma = [1 + \mathbf{P}\cdot\mathbf{P}/m_e^2 c^2]^{1/2}$ is the electron's relativistic factor. Using Eq. (2a) and setting $\lambda_m = L_d$, we find

$$\Delta\gamma(z) \simeq \frac{1}{4}\pi a_0^2 \delta^{-1} \frac{k_{p0}}{k_m}\left[J_0\left(\frac{1}{2}\delta k_m z + \frac{1}{2}\delta k_{p0}z_0\right) - J_0\left(\frac{1}{2}\delta k_{p0}z_0\right)\right], \quad (4)$$

where $z_0$ is the initial axial position of an electron. The energy gain increases with the laser amplitude through the larger wakefields driven by the pulse. As expected Eq. (4) limits to zero as $\delta \to 0$ or $k_m \to 0$: only the $n = 0$ spatial harmonic is present in this limit. The maximum acceleration will occur for electrons with initial axial positions near the peak of the $n = -1$ spatial harmonic, $\delta k_{p0}z_0 \sim -4$. The minimum value for $\delta$ is, however, limited: aside from experimental considerations such as density uniformity, the peak of $J_{-1}$ must occur within the length of the plasma channel, $L_{ch}$, such that $\delta$ can be no smaller than $\delta_{min} \sim 4/k_{p0}L_{ch}$. For $L_{ch} = 2$ cm and $n_0 = 7 \times 10^{18}$ $cm^{-3}$, $\delta_{min} = 4 \times 10^{-4}$ much smaller than the value of $\delta = 0.04$ used here. Setting $z_0 = -4/\delta k_{p0}$ and using the parameters described in the previous paragraph, Eq. (4) predicts an energy gain of $\Delta\gamma \sim 141$, whereas for the same parameters the dephasing limited energy gain of LWFA in a uniform plasma is $\Delta\gamma \sim 34$. Equation (4) underpredicts the energy gain observed in Fig. 1(c). This is somewhat surprising as the FWHM used for Fig. 1(c) is longer than the matched value used for deriving Eq. (4). As we will see, an enhancement in energy gain can result from the nonlinear evolution of the laser pulse.

Although quasi-phasematched LWFA permits energy gain over several dephasing lengths, the gain is eventually limited by electrons outrunning the structure of the spatial harmonic, the white curve in Fig. 1(c). One can show that for the $n = -1$ spatial harmonic, the length scale associated with this 'dephasing' process for an electron starting at $z_0 \sim -4/\delta k_{p0}$ with $\delta \gg \delta_{min}$ is $L_{-1} = 0.6\delta^{-1}L_d$. The inverse proportionality of $L_{-1}$ with $\delta$ results from the scale lengthening of the Bessel functions for small argument. For a modulation amplitude of $\delta = 0.04$, $L_{-1} = 15L_d$, an order of magnitude larger than standard LWFA. Based on $L_{-1}$, the maximum energy gain of QPM-LWFA is $\Delta\gamma_{Q,max} \simeq (1/8)[1 - J_0(2)]a_0^2\delta^{-1}k_{p0}L_d$, compared to the standard LWFA value of $\Delta\gamma_{S,max} = (\pi/16)a_0^2 k_{p0}L_d$, an increase of $2\pi^{-1}\delta^{-1}[1 - J_0(2)]$. For shorter acceleration distances, different initial positions, or smaller values of $\delta$, one can calculate the approximate energy gain directly from Eq. (4).

Figure 1(d) displays the modulation period required for phasematched acceleration by the $n = -1$ spatial harmonic and the theoretical energy gain after 2 cm as a function of plasma density for $\delta = 0.04$. The energy initially increases because the wakefield amplitude increases with plasma density. The decrease in energy results from the shortening of the maximum acceleration length due to the inverse density proportionality of $L_{-1}$: $k_p L_{-1} \propto n_0^{-1}$. We note that for larger densities, shorter pulse lengths would be required for satisfying the matching condition used when deriving Eq. (4).

Our estimate of the energy gain assumed that the electron's axial velocity was close enough to c that it did not undergo sufficient phase-slippage with respect to the $n=-1$ spatial harmonic. A condition on the minimum axial momentum for which this assumption is valid, or trapping condition, can be derived from the Hamiltonian of an electron interacting with the $n=-1$ spatial harmonic. Using $d\gamma/dt = -\omega_{p0}(v_z/c)E_z$ and defining $\Phi = k_{p0}v_g t - (k_{p0}+k_m)z$, the Hamiltonian takes the form

$$H = \frac{1}{8}\pi a_0^2 J_1(2)\sin(\Phi) - \left(\frac{v_g}{c}\right)(\gamma^2-1)^{1/2} + \left(1-\frac{k_m}{k_{p0}}\right)\gamma, \quad (5)$$

where $\Phi$ and $\gamma$ are the conjugate dynamic variables and the electron is assumed to be located near the peak of the spatial harmonic during the trapping process: $J_1[\delta k_{p0}(v_g t - z)/2] \sim J_1(2)$. Setting $k_m = 2\pi/L_d$ and using the fact that $H$ is a constant of motion, we find the threshold energy for trapping is $\gamma_{tr,Q} \simeq [4J_1(2)E_{max}]^{-1}$, where $E_{max} = \pi a_0^2/8$. For $a_0 = 0.25$, this predicts a trapping threshold of $\gamma_{tr,Q} \simeq 18$. For standard LWFA in the linear regime the trapping threshold is given by $\gamma_{tr,S}/\gamma_g \simeq (1+\gamma_g E_{max}) - [(1+\gamma_g E_{max})^2 - 1]^{1/2}$ where $\gamma_g = (1-v_g^2/c^2)^{-1/2}$ [21]. With the parameters defined earlier, this predicts $\gamma_{tr,S} \simeq 7$. The increased trapping threshold is a shortcoming of QPM-LWFA, but can be overcome with additional density tailoring to modify the plasma wave's phase velocity and other injection techniques [19,22-24].

To demonstrate quasi-phasematched laser wakefield acceleration, we conduct particle-in-cell (PIC) simulations using TurboWAVE [25]. TurboWAVE is a framework for solving the Maxwell-Lorentz system of equations for charged particle dynamics. The model is fully relativistic and fully electromagnetic. The fields are advanced with a Yee solver [26], and the sources are deposited using linear weighting, with charge conservation ensured by means of a Poisson solver.

The fields, particle trajectories, densities and currents were calculated on a 2D planar-Cartesian grid in a window moving at $c$. The window dimensions were 77 $\mu m$ × 438 $\mu m$ with 512 × 16384 cells in the transverse, $x$, and longitudinal directions, $z-ct$, respectively. The plasma density was ramped up over 200 $\mu m$. After the initial ramp, the plasma density followed $n_e(x,z) = n_0[1+\delta \sin(k_m z)](1+n_0'' x^2/2)$ with $n_0 = 7\times 10^{18}$ $cm^{-3}$, $n_0''$ set to guide a linearly polarized Gaussian mode with an $\exp(-1)$ field width $w_{ch} = 15$ $\mu m$, $\delta = 0.04$, and $\lambda_m = L_d = 5.0$ $mm$. To account for any non-ideal propagation effects, the value of $L_d$ was found numerically by first performing the simulation with $\delta = 0.0$.

The laser pulse was initialized with linear polarization in the x-direction, a $\sin^2$ temporal profile, and a Gaussian transverse profile. The pulse parameters were the same as above: $\lambda_0 = 800$ $nm$, $a_0 = 0.25$, $\sigma_{FWHM} = 30$ $fs$, and $w_{ch} = 15$ $\mu m$. We note that this is a pulse energy of only 14 $mJ$. The pulse started with its front edge at the beginning of the density ramp and centered along the channel axis. An electron bunch, also centered along the channel axis, was initialized with its center 200 $\mu m$ behind the peak of the laser pulse, a separation of $4/\delta k_{p0}$ as discussed above. The bunch had a transverse and longitudinal Gaussian profile with $\exp(-1)$ dimensions of 4 $\mu m$ and 8 $\mu m$ respectively. The peak bunch density was $n_b = 3.5\times 10^{16}$ $cm^{-3}$ with a total charge of 11 $pC$, parameters typical of LWFA experiments [27]. The simulations were conducted for bunch electrons with initial axial momentum of $P_z/m_e c = 30$ and $P_z/m_e c = 200$. We note that because the laser mode is channel guided and the pulse power, 0.5 $TW$, is lower than the critical power for self-focusing [28], $P_{cr} = 17(\omega/\omega_{p0})^2 GW = 4.2$ $TW$, differences between our 2D simulations and full 3D simulations should be minimal.

Figure 2 displays snapshots of background plasma density, laser intensity, and $P_z/m_e c = 30$ electron bunch density at propagation distances of 0.3 $mm$, 2.6 $mm$, and 5.2 $mm$. The laser pulse enters the plasma channel and excites a plasma wave, visible as the density oscillations trailing the pulse. As the bunch electrons enter the channel, they evolve in response to the wakefields. By 2.6 $mm$ the bunch electrons have been either laterally deflected or strongly focused. The transverse field of the $n=-1$ spatial harmonic is also quasi-phasematched to the electrons. For off-axis electrons, the quasi-phasematched transverse field provides either focusing or defocusing depending on the electron's initial longitudinal position. Comparing $E_z$ and $E_r$ in Eq. (2), we see that there are longitudinal regions of size $\lambda_p/4$ where electrons are both axially accelerated and focused. Electrons starting in these favorable regions remain on axis and continue to gain energy as they travel behind the laser pulse.

Comparisons of the maximum energy gain resulting from QPM-LWFA and standard LWFA are shown in Fig. 3. The solid and dashed black curves are the energy gain of electrons with an initial axial momentum of $P_z/m_ec=30$ in QPM-LWFA and LWFA respectively. The red curves are the energy gain of electrons with an initial axial momentum of $P_z/m_ec=200$. When the channel has a density modulation with period equal to the dephasing length, $\lambda_m=L_d$, electrons gain energy over several dephasing lengths. In a uniform channel, the electrons initially gain energy, but then lose energy as they outrun the accelerating phase of the wake. The energy oscillations in the modulated channel result from the electron being partially de-accelerated as it moves forward into the next phase of the plasma wave. After 1.5 $cm$, the energy gain reaches $\Delta\gamma \sim 104$ or $\Delta E \sim 51$ $MeV$ for both initial energies $\gamma_0=30$ and $\gamma_0=200$.

In spite of the pulse duration, $\sigma_{FWHM}=30$ $fs$, being longer than the matched duration, $\pi/ck_{p0}=21$ $fs$, the energy gain is nearly that predicted by Eq. (2): $\Delta\gamma \sim 106$. Thus, as with Fig. 1(c), the energy gain is larger than one would expect. Correspondingly, the gain rate increases near a distance of 6 $mm$ in Fig 3. The enhanced gain rate results from nonlinear temporal compression of the pulse. The pulse's ponderomotive force expels electrons from its path, forming a nonlinear electron density gradient within the pulse. The electron density decreases from the front of the pulse backwards, causing the middle and front of the pulse to undergo spectral redshifting. Because the group velocity decreases with decreasing frequency in plasma, the redshifted spectral components slide backwards, forming an optical shock near the back of the pulse [29]. The compression and corresponding increase in pulse intensity can be partly observed in Fig. 2. To demonstrate the effect more quantitatively, Fig. 4 shows the evolution of the pulse's on-axis temporal FWHM in black on the left vertical axis and on-axis intensity in red on the right vertical axis. Both quantities are normalized to their initial values. The decrease in FWHM is accompanied by a near equal increase in intensity. The product is essentially constant, suggesting that the pulse is not undergoing significant nonlinear focusing or spot oscillations due to unmatched guiding [30]. After 1.5 $cm$, the FWHM has reached less than the one third its original value. Likewise, the intensity has increased by nearly a factor of three. The nonlinear evolution of the laser pulse boosts the acceleration by compressing the initial unmatched pulse duration to a duration closer to the matched value $\sigma_{FWHM} \sim \lambda_p/2$, while, at the same time, increasing the wakefield amplitude through the increase in intensity. The inset in Fig. 4 displays the axial wakefield experienced by a relativistic electron in the corrugated channel. As a result of pulse compression, the axial field grows with axial distance. We note that this same process occurs in uniform plasmas, but for the parameters considered here, its utility for boosting acceleration is diminished due to dephasing.

We have investigated the application of quasi-phasematching enabled by modulated plasma channels to LWFA. Theory and self-consistent PIC simulations demonstrate that by setting the modulation period to the dephasing length, electron acceleration can occur over multiple dephasing lengths. This technique overcomes the dephasing limitation of laser wakefield acceleration. With millijoule level pulses, QPM-LWFA resulted in energy gains an order of magnitude higher than standard LWFA with identical parameters.

The authors would like to thank D. Gordon for the use of TurboWAVE and continued fruitful collaboration and P. Sprangle for useful discussions. This work was supported by DOE, DTRA, NSF, and ONR.

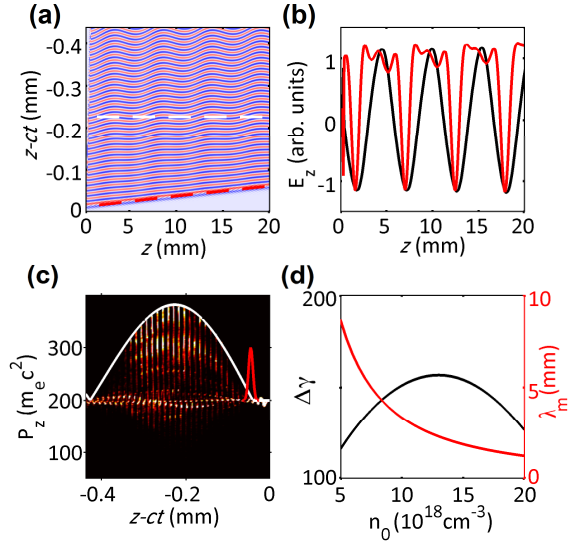

Figure 1. (a) On-axis wakefield in a modulated plasma as a function of speed of light frame coordinate, $z-ct$, and axial distance, $z$. The red dashed line marks the path taken by the on-axis peak of the laser pulse. (b) Axial wakefield experienced by an electron moving with an axial velocity near the speed of light at a position marked by the white dashed line in (a). The red and black lines are the fields experienced in a modulated and uniform plasma channel respectively. (c) Phase space density in the axial momentum-speed of light frame plane resulting after 2 mm of interaction. The white line shows the amplitude of the $n=-1$ spatial harmonic and the red line the envelope of the laser pulse. (d) Predicted energy gain, black, left vertical axis and matched modulation period red, right vertical scale as a function of average on-axis plasma density. Exact parameters are in the text.

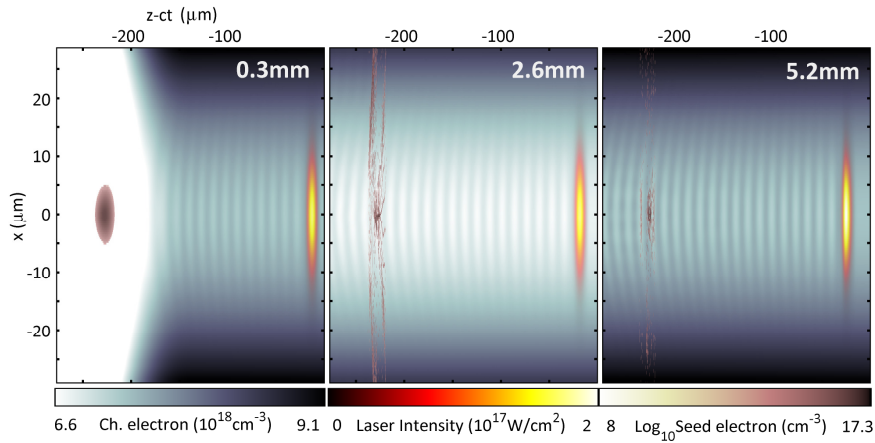

Figure 2. Background plasma density, laser pulse intensity, and electron bunch density as a function of tranverse position and speed of light frame coordinate at three axial distances: 0.3 $mm$, 2.6 $mm$, and 5.2 $mm$. The electron plasma wave is noticeable as the ripples in the background plasma density.

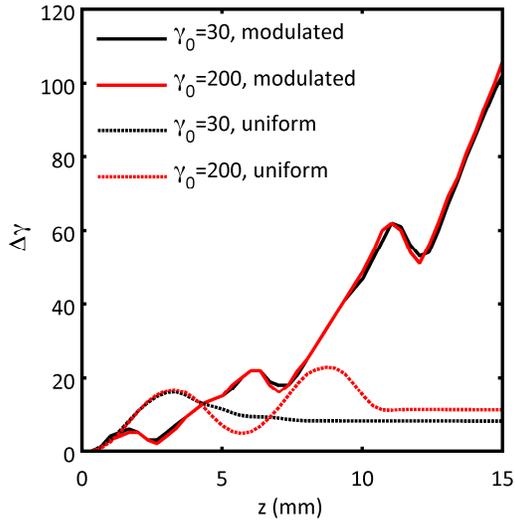

Figure. 3 Maximum energy gain as a function of distance for electrons accelerated in a modulated, solid, and uniform channel, dashed. The black and red lines are initial energies of $\gamma_0 = 30$ and $\gamma_0 = 200$ respectively.

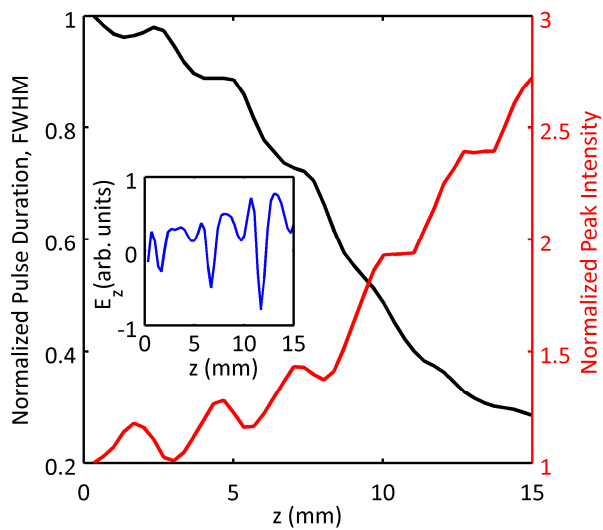

Figure. 4 On-axis temporal FWHM black, left vertical axis and peak intensity red, right vertical axis normalized to their initial values as a function of propagation distance. The inset displays the axial wakefield experienced by an electron moving near the speed of light. The wakefield amplitude increases due to the intensity increasing and the pulse duration shortening.